\documentclass{webofc}
\usepackage[varg]{txfonts}   
\usepackage{lineno}
\begin{document}

\title{Modification of heavy quark hadronization in high-multiplicity collisions at LHCb}
%
%

\author{\firstname{Chenxi} \lastname{Gu}\inst{1}\fnsep\thanks{\email{chgu@cern.ch}} on behalf of the LHCb collaboration 
}

\institute{Laboratoire Leprince-Ringuet, École Polytechnique
          }

\abstract{%
   The ratio of heavy flavor hadrons is very sensitive to the hadronization mechanism. This proceeding will present recent LHCb results on the cross-section ratios of $D_{s}^{+}/D^{+}$, $\Xi_{c}^{+}/\Lambda_{c}^{+}$ and $\Lambda_{b}^{0}/B^{0}$ in different collision systems. The significantly enhanced production ratios $D_{s}^{+}/D^{+}$ and $\Lambda_{b}^{0}/B^{0}$ with the increase of multiplicity may imply that hadronization mechanisms are modified in high-multiplicity events.
}
\maketitle
\section{Introduction}
\label{intro}
In the context of hadron colliders, heavy quarks primarily originate from hard parton-parton interactions in the initial stages of the collisions. These interactions are well described by perturbative QCD calculations, which rely on the factorization theorem. According to this theorem, the cross-sections of heavy-flavour hadrons depend on several key factors: the parton distribution functions within the incoming nucleons, the cross-section of hard parton-parton scattering, and the fragmentation functions. For different types of heavy-flavored hadrons, the contributions from the first two items are similar, and only the hadronization process makes the difference. Traditionally, assuming that hadronization of heavy quarks is a universal process independent of colliding systems, these fragmentation functions are parameterized using data collected from $ee$ or $ep$ collisions. Some recent measurements from the LHCb experiment on the $D_{s}^{+}/D^{+}$ and $\Lambda_{b}^{0}/B^{0}$ ratios have revealed a significant enhancement from low-multiplicity collisions to high-multiplicity collisions. These results indicate the existence of other hadronization mechanisms that are dependent on the collision size.

\section{The cross-sections ratio $D_{s}^{+}/D^{+}$ versus multiplicity in $p\mathrm{Pb}$ collisions}
\label{result1}
In $p\mathrm{Pb}$ collisions at $\sqrt{s_\mathrm{NN}} = 5.02$ TeV, LHCb conducted measurement of prompt $D_{s}^{+}$ and $D^{+}$ production, and studied relative production ratios~\cite{LHCb:2023kqs}. 
The $D_{s}^{+}/D^{+}$ ratio observed in backward rapidity is slightly higher than that in forward rapidity and $pp$ collisions~\cite{LHCb:2016ikn}. Previous findings from LHCb~\cite{LHCb:2021vww} suggest that backward rapidity result in a higher yield of charged particles compared to forward rapidity within the symmetric kinematic interval. These results suggest a potential enhancement in the $D_{s}^{+}/D^{+}$ ratio with increasing multiplicity.

\begin{figure}[h]
\centering
\includegraphics[width=12cm,clip]{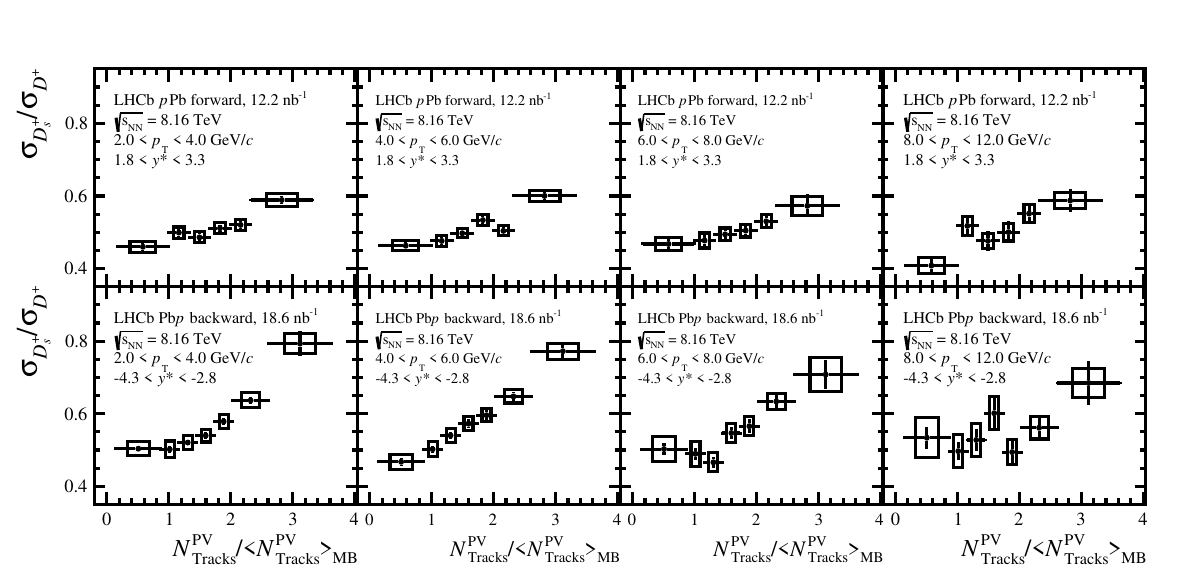}
\caption{Cross-section ratio $\sigma_{D_{s}^{+}}/\sigma_{D^{+}}$ versus the normalised event multiplicity in different $p_{\mathrm{T}}$ ranges for (top) forward and (bottom) backward rapidity~\cite{LHCb:2023rpm}.}
\label{fig-1}       
\end{figure}

Figure~\ref{fig-1} illustrates the cross-section ratio $D_{s}^{+}/D^{+}$ as a function of normalized event multiplicity in $p\mathrm{Pb}$ collisions at $\sqrt{s_\mathrm{NN}} = 8.16$ TeV~\cite{LHCb:2023rpm}. The event multiplicity is denoted by $N_{\mathrm{tracks}}^{\mathrm{PV}}$, representing the number of tracks used for reconstructing the primary vertex (PV). The charged particle multiplicity is normalized to the mean value observed in minimum-bias events. All panels exhibit a significant dependence on multiplicity, with a pronounced enhancement observed, particularly in the backward rapidity. This enhancement can be attributed to the combined effects of the coalescence mechanism and strangeness enhancement. However, this enhancement shows no tendency to weaken in the high transverse momentum ($6 < p_{\mathrm{T}} < 12$ GeV/$c$) ranges.

\section{Measurement of $\Xi_{c}^{+}$ production in $p\mathrm{Pb}$ collisions at $\sqrt{s_\mathrm{NN}} = 8.16$ TeV}
\label{result2}
\begin{figure}[h]
\centering
\includegraphics[width=5cm,clip]{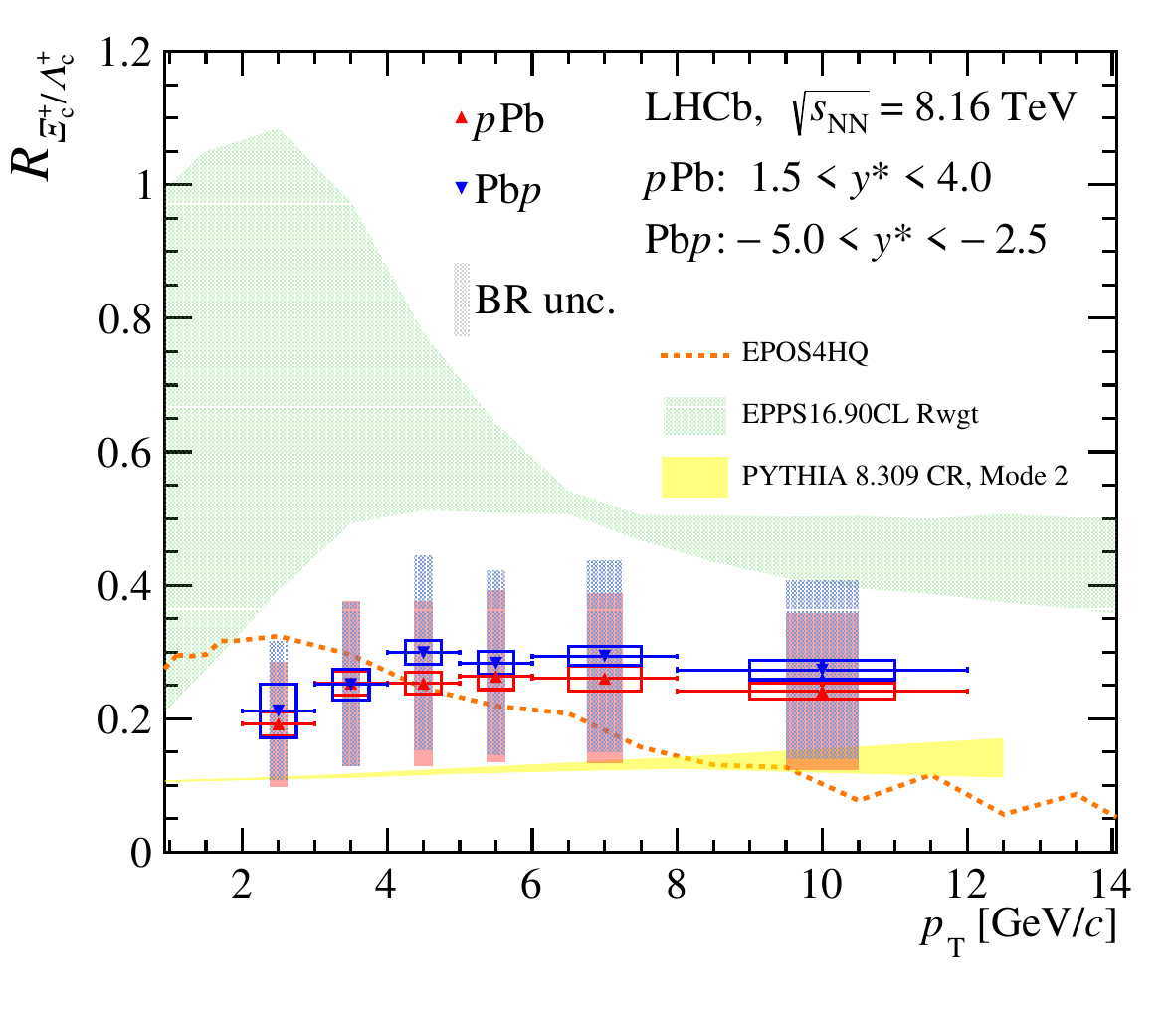}
\includegraphics[width=5cm,clip]{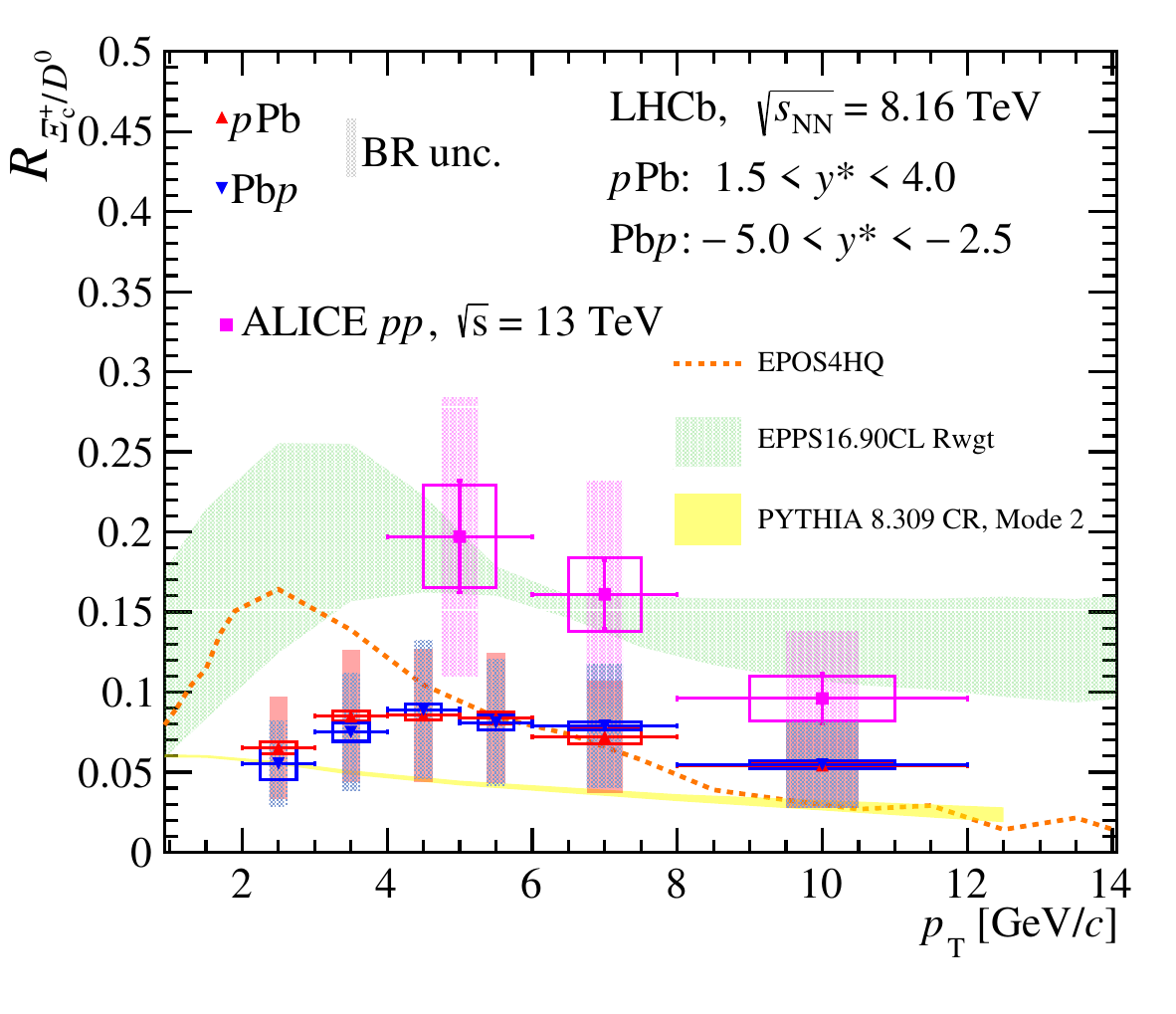}
\caption{Production ratios, (left) $\Xi_{c}^{+}/\Lambda_{c}^{+}$ and (right) $\Xi_{c}^{+}/D^{0}$, as a function of $p_{\mathrm{T}}$ in the $p\mathrm{Pb}$ (red triangles) and $\mathrm{Pb}p$ (blue triangles) data samples~\cite{LHCb:2023cwu}.}
\label{fig-2}       
\end{figure}

Figure~\ref{fig-2} shows production ratio of strange baryons to non-strange baryons, $\Xi_{c}^{+}/\Lambda_{c}^{+}$, and the production ratio of strange baryons to non-strange mesons, $\Xi_{c}^{+}/D^{0}$, in $p\mathrm{Pb}$ collisions at $\sqrt{s_\mathrm{NN}} = 8.16$ TeV~\cite{LHCb:2023cwu}. Both the $\Xi_{c}^{+}/\Lambda_{c}^{+}$ and $\Xi_{c}^{+}/D^{0}$ ratios show no significant $p_{\mathrm{T}}$ dependence. The $\Xi_{c}^{+}/\Lambda_{c}^{+}$ and $\Xi_{c}^{+}/D^{0}$ ratios are consistent within the error for forward and backward rapidity, but the $\Xi_{c}^{+}/\Lambda_{c}^{+}$ ratio in the backward rapidity is slightly higher than that in the forward rapidity.

The measurements are compared with the EPPS16 calculation~\cite{Eskola:2016oht}, but both $\Xi_{c}^{+}/\Lambda_{c}^{+}$ and $\Xi_{c}^{+}/D^{0}$ are overestimated. The calculations from Pythia 8.3 with color reconnection~\cite{Christiansen:2015yqa} and EPOS4HQ~\cite{Zhao:2023ucp, Werner:2023fne} are also shown in Figure~\ref{fig-2}, both of which are based on results from $pp$ collisions.

\section{Enhanced production of $\Lambda_{b}^{0}$ in high-multiplicity $pp$ collisions at $\sqrt{s} = 13$ TeV}
\label{result3}

\begin{figure}[h]
\centering
\includegraphics[width=11cm,clip]{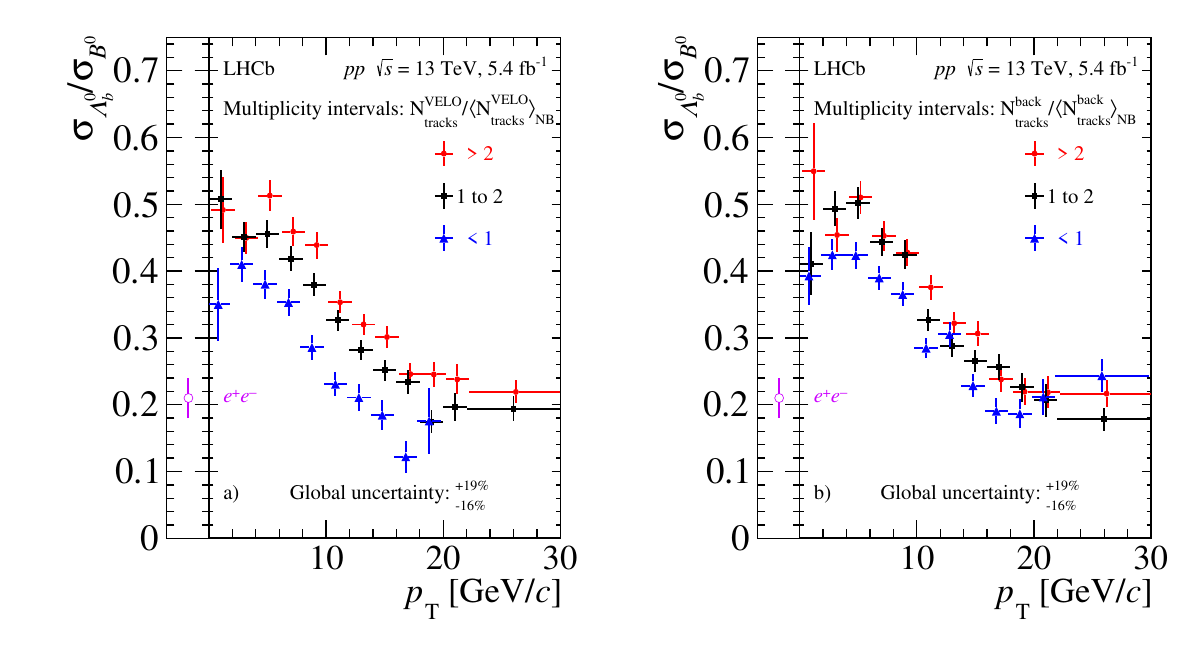}
\caption{Cross-section ratio $\sigma_{\Lambda_{b}^{0}}/\sigma_{B^{0}}$ as a function of $p_{\mathrm{T}}$ in different multiplicity bins.}
\label{fig-3}       
\end{figure}

Figure~\ref{fig-3} presents cross-sections ratio $\Lambda_{b}^{0}/B^{0}$ as a function of $p_{\mathrm{T}}$ in different multiplicity bins in $pp$ collisions at $\sqrt{s} = 13$ TeV~\cite{LHCb:2023wbo}. The event multiplicity is characterized by $\mathrm{N}_{\mathrm{tracks}}^{\mathrm{VELO}}$ and $\mathrm{N}_{\mathrm{tracks}}^{\mathrm{back}}$. 
$\mathrm{N}_{\mathrm{tracks}}^{\mathrm{VELO}}$ denotes the total number of charged tracks reconstructed in the VELO detector, while $\mathrm{N}_{\mathrm{tracks}}^{\mathrm{back}}$ represents the subset of $\mathrm{N}_{\mathrm{tracks}}^{\mathrm{VELO}}$ pointing away from the LHCb detector.
Both plots show a significant dependence on multiplicity, with clear enhancements observed, particularly evident when utilizing VELO tracks. At low $p_{\mathrm{T}}$, the $\Lambda_{b}^{0}/B^{0}$ ratio is significantly higher than the value observed in $e^{+}e^{-}$ collisions. However, as the $p_{\mathrm{T}}$ increases, the $\Lambda_{b}^{0}/B^{0}$ ratio tends to align with the results obtained from $e^{+}e^{-}$ collisions.

\begin{figure}[h]
\centering
\includegraphics[width=5cm,clip]{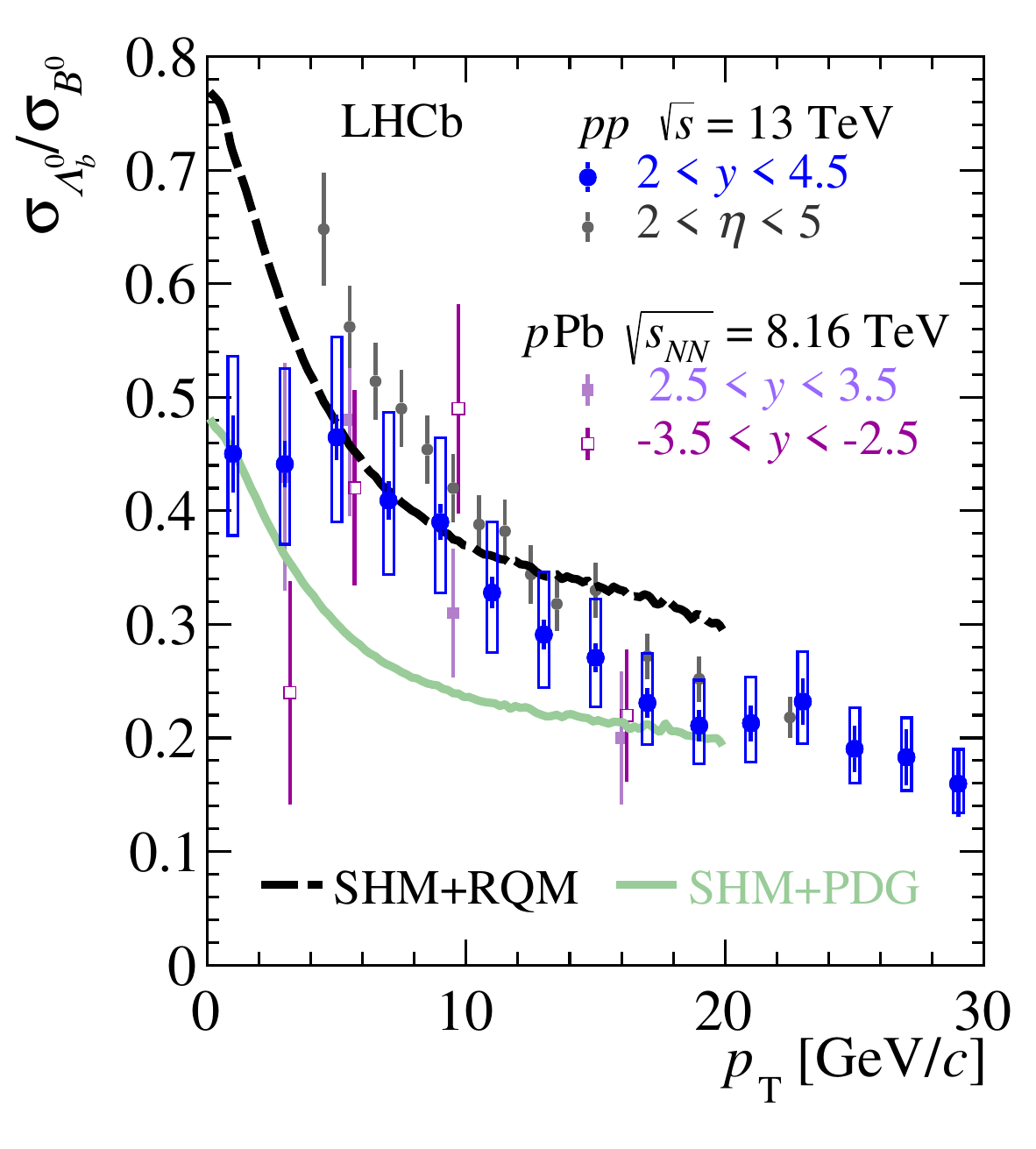}
\caption{Cross-section ratio $\sigma_{\Lambda_{b}^{0}}/\sigma_{B^{0}}$ as a function of $p_{\mathrm{T}}$.}
\label{fig-4}       
\end{figure}

Figure~\ref{fig-4} shows cross-sections ratio $\Lambda_{b}^{0}/B^{0}$ as a function of $p_{\mathrm{T}}$. The data are compared to previous $pp$ measurement~\cite{LHCb:2019fns} and $p$Pb measurement~\cite{LHCb:2019avm}, and generally consistent with them within uncertainties. Additionally, two curves from $b$ quarks statistical hadronization model~\cite{He:2022tod} are also shown in Figure~\ref{fig-4}. The green solid curve considers feeddown contributions from $b$ baryons which have been collected by Particle Data Group~\cite{ParticleDataGroup:2020ssz}. The black dashed curve takes into account feeddown contributions from an expanded set of $b$ baryons predicted by the Relativistic Quark Model~\cite{Ebert:2011kk}.

\section{Summary}
\label{summary}
In recent LHCb studies, cross-section ratios, $D_{s}^{+}/D^{+}$ and $\Lambda_{b}^{0}/B^{0}$, were measured in both low and high multiplicity collisions. $\Xi_{c}^{+}/\Lambda_{c}^{+}$ was measured in both $p$Pb and Pb$p$ collisions. The $\Xi_{c}^{+}/\Lambda_{c}^{+}$ shows little variation between forward and backward rapidity within uncertainties. The $D_{s}^{+}/D^{+}$ and $\Lambda_{b}^{0}/B^{0}$ show a significant enhancement in high-multiplicity collisions compared to low-multiplicity collisions. The $\Lambda_{b}^{0}/B^{0}$ ratio decreases with $p_{\mathrm{T}}$ and converges to the $e^{+}e^{-}$ result at high $p_{\mathrm{T}}$. These suggest the potential presence of other hadronization mechanisms in high-multiplicity collisions. Furthermore, it is also possible that in high-multiplicity collisions, the more contribution from excited state feeddown could lead to the observed enhancement in cross-section ratios. It is worth noting that this enhanced feeddown of SHM+RQM relative to SHM+PDG does not weaken as $p_{\mathrm{T}}$ increases, as predicted in Figure~\ref{fig-4}.

%
\bibliography{my-bib-database}
%
%
%
%

\section*{Acknowledgements}
This work has received funding from the European Union's Horizon 2020 research and innovation programme under the Marie Sk{\l}odowska-Curie grant agreement No.~899987 (EuroTechPostdoc2).

\end{document}